\pdfoutput=1
\documentclass[11pt]{arxiv-article}
	
\bibliography{References}

\usepackage{Definitions}

\usepackage{texdefs}

\usepackage{pifont}
\usepackage{hyperref}

\newcommand{\OfficialTitle}{
  An AdS/EFT correspondence \\
  at large charge
}
\title{\setstretch{1.4}
  {\color{Thoughtless}\Huge\textbf{\dosserif\OfficialTitle}}
}

\hypersetup{pdfauthor={Orestis Loukas and Domenico Orlando and Susanne Reffert and Debajyoti Sarkar},pdftitle={\OfficialTitle}}

\author{%
  \begin{minipage}{.97\linewidth}
    \vspace{1cm}
    \begin{center} \dosserif
      {\small
         \textbf{Orestis Loukas}\textsuperscript{\ding{72}},
         \textbf{Domenico~Orlando}\textsuperscript{\ding{72}\ding{73}},
         \textbf{Susanne~Reffert}\textsuperscript{\ding{72}} and
         \textbf{Debajyoti Sarkar}\textsuperscript{\ding{72}}}
    \end{center}
    \vspace{1cm}
    \authorBlock{\ding{72}}{\dosserif Albert Einstein Center for Fundamental Physics\\
      Institute for Theoretical Physics, University of Bern,\\
      Sidlerstrasse 5, CH-3012 Bern, Switzerland}
    \authorBlock{\ding{73}}{\dosserif INFN sezione di Torino | Arnold--Regge Center\\
      via Pietro Giuria 1, 10125 Turin, Italy}
  \end{minipage}
}

\date{}

\begin{document}

\setstretch{1.2}

\numberwithin{equation}{section}

\begin{titlepage}

  \newgeometry{top=23.1mm,bottom=46.1mm,left=34.6mm,right=34.6mm}

  \maketitle

  \thispagestyle{empty}

  \vfill\dosserif

  \abstract{\normalfont \noindent

    Considering theories in sectors of large global charge $Q$ results in a semiclassical \ac{eft} description for some strongly-coupled \acp{cft} with continuous global symmetries. Hence, when studying dualities at large charge, we can have control over the strongly coupled side of the duality and gain perturbative access to both dual pairs.
    
    In this work we discuss the AdS/CFT correspondence in the regime \(Q \gg C_T \gg 1\) where both the \ac{eft} and gravity descriptions are valid and stable ($C_T$ being the central charge).
    We present the observation that the ground state energy as a function of the Abelian charge $Q$ for a simple \ac{eft} in some three-dimensional \acp{cft} coincides with the expression for the mass of an \acl{adsrn} black hole as a function of its charge.
    Using this observation we propose a tentative dictionary relating \ac{cft}, \ac{eft} and holographic descriptions.
    We also find agreement for the higher-derivative corrections on both sides, suggesting a large-$C_T$ expansion on the \ac{eft} side.
  }

\vfill

\end{titlepage}

\restoregeometry

\setstretch{1.2}

\tableofcontents

\section{Introduction}
\label{sec:introduction}

\acresetall
\acused{adscft}
\acused{adseft}

Strongly coupled \acp{cft} play an important role in theoretical physics as they appear in many contexts, such as at fixed points of renormalization group flows and in the description of critical phenomena. In most cases, they however elude analytic treatments. A possible approach consists in considering such models in sectors of fixed and large global charge which gives rise to important simplifications~\cite{Hellerman:2015nra, Alvarez-Gaume:2016vff, Loukas:2016ckj, Loukas:2017lof, Loukas:2017hic, Hellerman:2017efx, Hellerman:2017veg, Hellerman:2017sur, Monin:2016jmo, Jafferis:2017zna, Bourget:2018obm,Hellerman:2018xpi}.\footnote{The conformal bootstrap often provides a viable approach, see~\cite{Rattazzi:2008pe,El-Showk:2014dwa,ElShowk:2012ht,Rychkov:2016iqz,Simmons-Duffin:2016gjk} and references therein.} It enables us to write an effective action for the fluctuations around the lowest-energy state at fixed charge $Q$ (which breaks time-translation invariance) which are in general encoded by one or more Goldstone bosons.
Most terms in this effective action are suppressed by negative powers of the large charge. 

The effect of working at large charge is a \emph{classicalization} of the physics, in the sense that we can use a semiclassical description and a perturbative expansion for a strongly coupled system. We expect this classicalization of strongly coupled systems at large quantum number to have profound implications for  duality, which usually bridges two different theories; one at strong coupling and the other at weak coupling. Since the above argument implies classicalization of a theory \emph{both} at strong and at weak coupling, it implies the existence of a regime of overlap where both dual descriptions are simultaneously valid. In other words, the large quantum number expansion makes nontrivial strong-weak coupling dualities manifest by simultaneously classicalizing both sides. This approach was first applied in~\cite{Hellerman:2015nra} evaluating a duality~\cite{Banks:1977cc,Peskin:1977kp} between the effective actions for the $O(2$) model and for an Abelian gauge theory at large charge. This duality was further exploited in~\cite{Cuomo:2017vzg}, allowing the authors to study operators of non-vanishing spin and large charge. 

An important duality which involves on one side a strongly coupled \ac{cft}, is the well-known \ac{adscft} correspondence~\cite{Maldacena:1997re,Witten:1998qj,Gubser:1998bc}. In our present context, having a certain amount of control over the strongly coupled \ac{cft} invites the question of whether we can apply the \ac{adscft} correspondence successfully in such a case. Although AdS$_{d+1}$/CFT$_d$ was originally discovered in the context of $\mathcal{N}=4$ \ac{sym} theory and \tIIB superstrings for $d=4$, it is now better understood as providing a more general `strong-weak' duality between an AdS gravity theory in $(d+1)$ spacetime dimensions and a gauge theory living at the conformal boundary of AdS (for a classic review, see e.g.~\cite{Aharony:1999ti}).

In this paper, we will confine ourselves to the low-energy or small string length $l_s=\sqrt{\alpha'}$ limit on the gravity side.
Due to the standard \ac{adscft} identification ${L}/{l_s}\approx\lambda_t^{p}\gg 1$ ($L$ being the AdS radius and $p$ some positive power depending on the precise nature of the duality) this implies that the dual \ac{cft} originally (before considering it at large charge $Q$) has a strong coupling parameter $\lambda_t$. We will also consider small Newton's constant $G$, which means that our gravity sector consists of the ordinary Einstein--Hilbert action plus additional fields. Usually, a small $G$ on the gravity side implies \acp{cft} at large central charge $C_T$ by the relation\footnote{The proportionality factor is a constant, which depends on dimensions and conventions.}
\begin{equation}
\frac{L^{d-1}}{G}\propto C_T\gg 1.
\end{equation}
Here $G$ is the $(d+1)$-dimensional Newton's constant, which is also related to the $d+1$ dimensional Planck length $l_p$ and Planck mass $m_p$  as $G=\left(l_p\right)^{d-1}=\left({m_p}\right)^{-(d-1)}$. In what follows, we will be working with four dimensional ($d=3$) Planck length and Planck mass. For a dual, boundary matrix model with $SU(N)$ gauge group, the central charge $C_T$ is usually proportional to the total number of degrees of freedom, $C_T = \order{N^q}$. For matrix models,  \(q = 2\). In these cases, with slight abuse of language, the $G$ corrections in the bulk are often interpreted as $1/N$ corrections at the boundary.
We will discuss this further in the following.

\medskip

The main result of this paper is that the ground state energy at fixed charge\footnote{This is mapped via the state-operator correspondence to the conformal dimension of the lowest-lying operator of charge $Q$ in the \ac{cft}.} in our \ac{eft} precisely reproduces the mass of the non-supersymmetric extremal \ac{adsrn} black hole, to all orders in $Q$. 
In the main part of the paper, we will identify the central charge factor $C_T$ in the context of the boundary theories we are interested in.
We will also include finite (but small) $G$ corrections in terms of higher-derivative terms in the metric~\cite{Yale:2010jy} in a controlled fashion and understand their implications for the \ac{eft}. These corrections imply that at the boundary, the central charge is always large but finite.
Usually in the above limit, one has a classical, weakly coupled gravity theory in the bulk and a strongly coupled \ac{cft} at the boundary.
However, as mentioned above, if we consider both sides of the correspondence at large charge, we are in a novel sector of the duality where both sides are weakly coupled.

Our observation about the precise identification of the functional forms of the energies for the {\ac{eft}} and the extremal {\ac{adsrn}} black hole is quite tantalizing, but leaves us with a number of questions about the identification of a precise holographic dual to the large-charge {\ac{eft}}.
  One first observation is that the {\ac{adsrn}} black hole has finite entropy, while the {\ac{eft}} is a simple gapped theory to which it is difficult to associate a zero-temperature entropy.
  Another issue is related to the existence of quasinormal modes with a non-vanishing (imaginary) sound attenuation constant~\cite{Edalati:2009bi}.
  While at leading order in \(Q\), the spectrum of the black hole coincides with the one of the {\ac{eft}} (as we discuss in Section~\ref{sec:quantum-fluctuations}), the latter does not have imaginary components in the subleading terms~\cite{Alvarez-Gaume:2016vff}.
In spite of these issues that we intend to address in future work, we believe that the all-order correspondence in the form of the energies is an interesting and unexpected result that deserves further investigation.

\medskip

The plan of this paper is as follows. In Section~\ref{sec:linear-sigma-model-and-black-hole}, we first discuss the effective action for a \ac{cft} at large global $U(1)$ charge (Sec.~\ref{sec:effective-action-cft}) and then the \ac{adsrn} black hole in Einstein--Hilbert gravity (Sec.~\ref{sec:reissn-nordstr-BH}).
We find an identical functional dependence of the energy in the \ac{eft} and of the mass of the black hole as functions of the charge, which we use to propose a tentative \ac{adseft} dictionary (Sec.~\ref{sec:dictionary}). In Section~\ref{sec:higher-order-terms}, we discuss the effects of higher-order terms both on the \ac{eft} and the gravity side of the correspondence.
In Section~\ref{sec:quantum-fluctuations}, we review the leading-order quantum corrections from the point of view of the gravity dual as given in~\cite{Edalati:2010pn,Edalati:2010hk} which agree, at leading order in the charge, with our \ac{eft} results.
In Section~\ref{sec:discussion}, we conclude with a discussion of our findings and give further directions.
In Appendix~\ref{sec:app}, we briefly discuss the free energy of the boundary theory.

\section{The linear sigma model and the black hole}
\label{sec:linear-sigma-model-and-black-hole}

In the following, we will present the observation that we can match the expression for the ground state energy of an \ac{eft} to the expression for the mass of a %
four-dimensional \ac{adsrn} black hole. 
We consider the lowest-energy state of a three--dimensional \ac{cft} with a global symmetry in a sector of fixed and large  charge associated to a $U(1)$ of the global symmetry.
For the \ac{eft} describing the low-energy physics at large charge, it is natural to investigate the corresponding thermodynamic quantities of an extremal \ac{adsrn}, as was first anticipated by~\cite{Nakayama:2015hga} and later by~\cite{Jafferis:2017zna}.  
We will briefly comment on this issue again in the conclusions, but for now, leave a detailed study of the inclusion of thermodynamic effects on both sides of the correspondence for future work. Nonetheless, comparing the two expressions and invoking the \ac{adscft} correspondence allows us to find a precise dictionary between the \ac{eft} and the black hole physics.

\subsection{Effective action for a CFT at large global $U(1)$ charge}
\label{sec:effective-action-cft}

We start by introducing an \ac{eft} for a three-dimensional \ac{cft} with an Abelian global symmetry in the limit of large charge \(Q\).
The idea is to restrict ourselves to a subsector of the theory where the charge is fixed and to use the inverse charge \(1/Q\) as a controlling parameter for a perturbative expansion.
The \ac{cft} is generic so there are a priori no small parameters apart from \(1/Q\). 

We assume that the ground state does not spontaneously break translational invariance. This means that we will expand around a classical fixed-charge ground state which is homogeneous in space.
This is consistent with previous results for the \(O(N)\) vector models~\cite{Hellerman:2015nra, Alvarez-Gaume:2016vff} and leads to very accurate predictions for the conformal dimensions of the lowest operator at fixed charge, which for the  \(O(2)\) model were verified via lattice calculations~\cite{Banerjee:2017fcx}.
The theory lives on a manifold with a typical scale \(r_0\).
{Fixing a global charge \(Q\) induces a scale related to the charge density \(\approx Q/r_0^2\).
As discussed in~\cite{Hellerman:2015nra}, this allows us to realize a consistent \ac{eft} with a cutoff  \(\Lambda_{\acs{eft}}\) that satisfies the hierarchy}
\begin{equation}
\label{eq:EFT_ValidityRegime}
  \frac{1}{r_0} \ll \Lambda_{\acs{eft}} \ll \frac{Q^{1/2}}{r_0}.
\end{equation}
In this regime, the system is described by an action that is approximately scale-invariant.
In the limit of large Abelian charge \(Q\) we assume that the dynamics is controlled by a single Goldstone field \(\chi\).
This is the situation for the \(O(N)\) vector models~\cite{Alvarez-Gaume:2016vff} and for some models with adjoint-valued order parameter~\cite{Loukas:2017lof, Loukas:2017hic}.
We will come back later to the issue of the presence of other gapless modes.

A simple Lagrangian {density} that realizes this physics within the regime of validity of Eq.~\eqref{eq:EFT_ValidityRegime} is given by
\begin{equation}
  \label{MatrixModels:IRaction}
  \mathcal L_{\acs{ir}} = \frac{1}{2} \Tr[\del_\mu \Phi \del^\mu \Phi] - \frac{\mathcal{R}}{16} \Tr[\Phi^2] - P_6(\Phi),
\end{equation}
where \(\Phi\) is a field that transforms in the adjoint of \(SU(N)\),  \(P_6 (\Phi)\) is an \(SU(N)\)-invariant sixth-degree polynomial and $\mathcal R$ is the Ricci scalar of the manifold \(\setR_t \times \mathcal{M}\) on which the theory lives.
Eventually, to compute the anomalous dimension of operators we will choose 
$\mathcal{M} = S^2(r_0)$, 
so that $\mathcal R=2/r_0^2$\,.
In Section~\ref{sec:higher-order-comparison} we will see that this \textsc{ir} Lagrangian is indeed a good description of our physics when $N$ is large.
Taking \(\Phi\) to have classical mass dimension \(1/2\), this model is manifestly (classically) scale invariant.
Furthermore, the action is by construction invariant under global \(SU(N)\) transformations,
\begin{equation}
V \in SU(N):\quad \Phi ~\rightarrow~ Ad[V] \Phi = V \Phi V^{-1}
,
\end{equation}
and the corresponding Noether current is given by the commutator
\begin{equation}
J_\mu = i b \comm{\Phi}{\del_\mu \Phi},
\end{equation} 
where $b$ is an order-one parameter that depends on the global properties of the matrix-valued field \(\Phi\) (see~\cite{Loukas:2017lof}).

For the purposes of this paper, we choose to fix the \(U(1)\) charge corresponding to the Chevalley generator 
\begin{equation*}
H_1 = \diag(1,-1, \underbrace{0,\dots,0}_{N-2})
\end{equation*}
and to look for the corresponding homogeneous solution to the \ac{eom}.
This is equivalent to determining the state of lowest energy $E$ at finite charge density $J_0$.
Up to obvious global symmetries, the desired solution takes the form
\begin{equation}
\label{MatrixModel:HomogeneousSolution}
  \Phi(t) = \Ad[{e^{i\mu t h}}] \Phi_0~,
\end{equation}
where
the direction of the time-dependent \textsc{vev} $e^{i\mu t h}$ can be embedded within the maximal torus as
\begin{equation}
  2h =
  \begin{cases}
    \diag(1,-1,...,1,-1) & \text{if \(N\) is even} \\
  \diag(1,-1,...,1,-1,0) & \text{if \(N\) is odd}
  \end{cases}
\end{equation}
and the time-independent \textsc{vev} $\Phi_0$ is a Hermitian matrix with only two non-vanishing elements $(\Phi_0)_{12}=(\Phi_0)_{21} =v/\sqrt{2}$.
The chemical potential $\mu$ and the radial amplitude $v$ are related via the \ac{eom}:
\begin{equation}
\label{CFT:classicalEOM}
\mu^2 = \lambda v^4 + \frac{\mathcal R}{8},
\end{equation} 
where $\lambda$ is a (dimensionless) Wilsonian parameter of order one that depends on the precise form of the scalar potential \(P_6 \).
The Noether current for such a ground state is given by 
\begin{equation}
\label{MatrixModels:U1current}
J_0 = b \mu v^2\, H_1%
\qand
\vec J = \vec 0
,
\end{equation}
so that the corresponding \(U(1)\) charge (henceforth denoted by $Q$) is given by
\begin{equation}
\label{O2:U1charge}
Q\, H_1 = \int_{\mathcal{M}} \dd[2]{x} J_0 = \vol(\mathcal{M}) b \mu v^2\, H_1
~.
\end{equation}
Notice that large charge implies via Eq.~\eqref{CFT:classicalEOM} a large chemical potential, $\mu=\order{\sqrt Q}$.

In the case of a two-sphere of radius $r_0$, \emph{i.e.} \(\mathcal{M} = S^2(r_0)\),
by taking the \ac{vev} of the Hamiltonian associated to Eq.~\eqref{MatrixModels:IRaction}
we can easily write the energy of the ground state as a function of the total charge $Q$: %
\begin{equation}
  \label{eq:EFT-energy}
  Er_0  = E(Q)r_0  = \frac{\pi}{6\sqrt 2} \frac{1}{\sqrt\lambda} \pqty{ 2 + \sqrt{ 1 + \frac{4 \lambda Q^2} {b^2 \pi^2} } } \sqrt{ \sqrt{ 1 + \frac{4 \lambda Q^2}{b^2 \pi^2}} - 1 }.
\end{equation}
The form of the ground-state energy dependence on the charge in Eq.~\eqref{eq:EFT-energy} is quite general and applies to any system where the type of symmetry-breaking pattern is dictated by such a time-dependent \ac{vev}.
In addition to the matrix model we have reviewed here, it was found to describe the \(O(2)\) vector model and the supersymmetric \(\mathcal{N} = 2 \) chiral model with superpotential \(W = \Phi^3\) in~\cite{Hellerman:2015nra} and the \(O(N)\) vector model in~\cite{Alvarez-Gaume:2016vff}.
{Note in particular that the form of Eq.~\eqref{eq:EFT-energy} does not depend on the details of the global symmetry group and it remains the same for all models with at least a \(U(1)\) global symmetry.}
We will see in the next section that \(E = E(Q)\) has precisely the same functional dependence as the mass of a \ac{adsrn} black hole as function of its charge.

Of course the \ac{ir} action associated to Eq.~\eqref{MatrixModels:IRaction} is just a truncation.
The actual Wilsonian action contains infinitely many higher-derivative terms which are compatible with the symmetries of the system.
In~\cite{Hellerman:2015nra} it was first observed that when expanding around the fixed-charge ground state, all these terms are controlled by powers of the inverse charge \(1/Q\) and do not change the effective description.
We will come back to their role and see how to understand the validity of the Lagrangian in Eq.~\eqref{MatrixModels:IRaction} from the point of view of the holographic correspondence in Section~\ref{sec:higher-order-terms}.

Another source of corrections are quantum effects.
The analysis of the fluctuations around the homogeneous ground state  $\Phi(t)$  
shows that the low-energy dynamics corresponds to the spontaneous symmetry breaking pattern~\cite{Loukas:2017lof,Loukas:2017hic}:
\begin{equation}
  U(2\floor{N/2}-1) \to  U(2\floor{N/2}- 2) .
\end{equation}
For a generic system, its physics is described by one relativistic Goldstone $\pi =\pi(t,\vec x)$ 
with dispersion relation in momentum space ($p\equiv\vert\vec p\vert$)
\begin{align}
\label{FirstSpeedOfSound}
  \omega_\text{rel}(p) = c_s p = \frac{1}{\sqrt 2} p
\end{align}
and \((2\floor{N/2}-2)\) non-relativistic gapless modes with Galilean dispersion relation
\begin{equation}
  \label{eq:non-relativistic-Goldstones}
  \omega_\text{n.r.}(p) = \frac{p^2}{2\mu} .
\end{equation}
In general, the existence of gapless modes around the homogeneous solution Eq.~\eqref{MatrixModel:HomogeneousSolution} is expected to modify the ground-state prediction for the energy on $S^2(r_0)$ in Eq.~\eqref{eq:EFT-energy}.
However, non-relativistic modes do not contribute to the Casimir energy on $S^2(r_0)$ by construction \cite{Alvarez-Gaume:2016vff} and
their effect appears at order $1/\sqrt Q$ and below. 
The first quantum correction to the energy formula is of order \(Q^0\) and comes from the Casimir energy on $S^2(r_0)$ of the leading Goldstone field that we compute in Section~\ref{sec:quantum-fluctuations}.
{Just like the classical energy, this contribution is universal for all generic systems with at least a \(U(1)\) global symmetry.}
We will see in Section~\ref{sec:quantum-fluctuations} how to recover it when studying the fluctuations of metric and Maxwell field in the bulk around the dual \ac{adsrn} background.

\subsection{The anti-de Sitter--Reissner--Nordstr\"{o}m black hole in Einstein--Hilbert gravity}
\label{sec:reissn-nordstr-BH}

We have seen that both the classical ground state in the \ac{eft} and the first quantum correction depend only on the existence of a \(U(1)\) symmetry.%
\footnote{In fact, in the case of matrix models it is possible to independently fix more than one $U(1)$ charge for a generic potential $P_6$, as shown in~\cite{Loukas:2017hic}. It is suggestive to think such multi-charge solutions might correspond to multi-center \ac{adsrn} black holes. For now, we leave this topic for future investigation.} 
On the gravity side, we can try to reproduce this physics starting with a $U(1)$ gauge theory coupled to Einstein--Hilbert gravity in four dimensions with action
\begin{equation}
  \label{EHM:Action}
  S %
  = \frac{1}{16\pi G} \int \dd[4]{x} \sqrt{-g} \bqty{ \mathcal R + \frac{6}{L^2}} - \frac{1}{4 e^2} \int \dd[4]{x} \sqrt{-g} \, F_{AB}F^{AB}
\end{equation}
with negative cosmological constant \(-6/L^2\).
The electric coupling constant is denoted by $e$ and \(G\) is Newton's constant.
Since there is no natural reason to have a hierarchy between the gravity and the gauge sectors, we expect the respective couplings $L^2/G$ and $1/e^2$ to be of the same order.
On general grounds~\cite{DUFF1983254}, this action will be a good approximation of the physics when the loop-counting parameter is small, \emph{i.e.} when
\begin{equation}
  \label{eq:EHM-action-consistency}
  \frac{L^2}{G} \approx \frac{1}{e^2} \gg 1 .
\end{equation}
We will come back to this fact in the next section when we introduce the holographic dictionary and then we will discuss the higher-order corrections in detail in Section~\ref{sec:higher-order-terms}.

We consider the most symmetric classical solution to the \ac{eom}, an \acl{adsrn} black hole with charge
 $Q$ and \ac{adm} mass $M$. In our conventions the (semiclassically integer-quantized) charge of a point particle at the origin is defined through
  \begin{equation}
    Q = \frac{1}{e^2} \int_{S^2_\infty} *F = \frac{4 \pi}{e^2} \lim_{r \to \infty} r^2 F_{tr}
    ~,    
  \end{equation}
  where \(S^2_\infty\) is the boundary of a constant-time slice at infinity.
The corresponding metric (asymptotically AdS as $r\rightarrow\infty$) and gauge field are given by
\begin{align}
\label{eq:RNsolution}
  \dd{ s^2} &= %
  -f(r)\dd{t^2} + \frac{\dd{r^2}}{f(r)}+ 
  r^2\left(\dd{\theta^2} + \sin^2 \theta \dd{\phi^2}\right)
  =
  - f(r)\dd{ t^2} + \frac{\dd{ r^2}}{f(r)}+ r^2 \dd{\gO}~,
  \nonumber
  \\
  A_t &= \frac{e^2Q}{4\pi} \left(\frac{1}{r} - \frac{1}{r_h} \right),
\quad A_r=0 ~\text{ and }~\vec A = 0,
\end{align}
where the usual \ac{adsrn} blackening factor reads %
\begin{equation}
	\label{AdSMetricFunction}
f(r) =
1 - \frac{2G}{r}M %
+\frac{e^2G}{4\pi r^2}Q^2
+\frac{r^2}{L^2}
~.
\end{equation}
The holographic coordinate runs from the boundary $r=\infty$ to the horizon $r=r_h$. 
The boundary geometry is conformally equivalent to $\mathbb R_t \times S^2(L)$. 
The chemical potential\footnote{Formally, one can switch from the chemical potential $\mu$ to the charge density $J^0$ by Legendre-transforming~\cite{Loukas:2016ckj}. Equivalently, one can twist the boundary conditions for the gauge field $A_M$~\cite{Herzog:2008he}.}
of the boundary \textsc{cft}
is then introduced as  
\begin{align}
	\label{eq:Dictionary:ChemicalPotentil_RadialVEV}
	\mu = \lim_{r \to \infty}A_t = -\frac{e^2}{4\pi r_h} Q
	\,.
\end{align}
In fact, this $\mu$ will ultimately turn out to be the one parametrizing the classical solution Eq.~\eqref{CFT:classicalEOM} on the \ac{eft} side.

In order to set up the dictionary, we want to work in the extremal limit, i.e.\ at zero temperature, to relate to our previous \textsc{eft} description.
This is achieved by looking for a double zero of $f(r)$, \emph{i.e.} the solution to the equations
$
f(r_h) = f'(r_h) = 0.
$

First, the horizon is always determined by the condition
\begin{equation}
	\label{BlackHoleMass}
f(r_h) \overset{!}{=}0
\quads{\Rightarrow}
2 G M = %
r_h +\frac{r_h^3}{L^2}+\frac{e^2G}{4\pi r_h}Q^2.    
\end{equation}
This relates the \ac{adm} mass of the black hole to the position of the horizon and to $Q$.
Plugging the expression of  $M$ back into~\eqref{AdSMetricFunction} we find
\begin{equation}
  \label{NoHair:MetricFunction2}
  f(r) = 
  1-\frac{r_h}{r}
  + \frac{e^2G}{4\pi r^2} \left(1-\frac{r}{r_h}\right)Q^2 + \frac{r^2}{L^2} \left(1-\frac{r_h^3}{r^3}\right)
  ~.
\end{equation}
The other condition to be satisfied for extremality is
\begin{equation}
	\label{HawkingTemperature}
0\overset{!}{=}
 f^\prime (r_h)
=
- \frac{e^2G}{4\pi r_h^3}Q^2 + \frac{1}{r_h} + \frac{3r_h}{L^2}
=
\frac{M'(r_h)}{r_h}
~,
\end{equation}
resulting in
\begin{equation}
r_h =  \frac{L}{\sqrt6} \sqrt{\sqrt{\frac{3}{\pi}\frac{G}{L^2}e^2Q^2+1}-1}
~.
\end{equation}
From the last part in Eq.~\eqref{HawkingTemperature} we see that taking $T=0$ implies a charged black hole of the smallest possible mass $M$. This is reminiscent of the lowest-energy condition leading to Eq.\ ~\eqref{eq:EFT-energy} on the \textsc{eft} side.
The mass of the extremal black hole is ultimately given by
\begin{equation}
  \label{eq:BH-mass}
   M L = 
   \frac{1}{3 \sqrt{6}} \frac{L^2}{G} 
   \pqty{2 + \sqrt{1 + \frac{3}{\pi}\frac{G}{L^2}e^2Q^2} }
   \sqrt{\sqrt{1 + \frac{3}{\pi}\frac{G}{L^2}e^2Q^2 } - 1 } 
   ~,
\end{equation}
expressed only in terms of the charge $Q$ and the bulk parameters $L,G$ and $e$.
This expression has precisely the same functional form as the energy on the \ac{eft} side~\eqref{eq:EFT-energy}.
If we identify $M$ with the energy on the \ac{cft} side, the extremality condition in the bulk is nothing but the lowest-energy condition (variational problem for spin-0 fields) in the \ac{eft}.
  One crucial observation, which goes beyond the correspondence with the \ac{eft}, is that the mass--charge relation in the black hole admits a large-charge expansion, which means that at least the philosophy at the basis of the work in~\cite{Hellerman:2015nra} can be applied.

Since we want to compare this result to the \ac{eft} of the previous section, we would like to think of \(M=M(Q)\) as an expansion for \(Q \gg 1\).
The expression in Eq.~\eqref{eq:BH-mass} suggests that the natural ``large parameter'' for the expansion is a combination of the charge $Q$ and the rest of the bulk parameters $G$, $L$ and $e$
so that \(M(Q)\) is expanded as
\begin{equation}\label{bhmass-Q}
  M L = 
  \frac{1}{3 \sqrt{6}} \frac{L^2}{G} \bqty{ \pqty{\sqrt{\frac{3}{\pi}} \frac{\sqrt{G}}{L} e Q}^{3/2} + \frac{3}{2} \pqty{\sqrt{\frac{3}{\pi}} \frac{\sqrt{G}}{L} e Q}^{1/2}} + \order{\pqty{\sqrt{\frac{3}{\pi}} \frac{\sqrt{G}}{L} e Q}^{-1/2}} .
\end{equation}
From this point of view, the fact that the \ac{eft} energy in Eq.~\eqref{eq:EFT-energy} agrees with the black hole mass can be seen as a matching of infinite terms in the asymptotic expansion at large charge.

At the beginning of this section we have seen that the classical gravity description is consistent in the regime of Eq.~\eqref{eq:EHM-action-consistency}.
In terms of the large parameter that we have found in the expansion, this means
\begin{equation}
  \label{Qerel}
  \sqrt{\frac{3}{\pi}} \frac{\sqrt{G}}{L} e Q \gg 1 ~\implies~ Q \gg \frac{L}{\sqrt{G}} \frac{1}{e} \gg 1 .
\end{equation}
We will see in the next section, by invoking 
our \ac{adseft} dictionary, that this means that the expansion is well-defined when 
$Q\gg\sqrt{C_T C_J}\gg1$.
\subsection{AdS/EFT/CFT dictionary}
\label{sec:dictionary}

The identical functional dependence of the energy in the \ac{eft} in Eq.~\eqref{eq:EFT-energy} and of the mass of the black hole as functions of the charge in Eq.~\eqref{eq:BH-mass} suggests that we can match these two quantities and propose a dictionary between the \ac{eft} description and the dual gravity system.
  As much as this might seem only partial evidence for the identification of a precise holographic duality, it is interesting to explore its possible consequences. 

For starters, remember that at the boundary, the geometry is conformally equivalent to \(\setR_t \times S^2(L)\).
This is where the \ac{cft} on \(S^2(r_0)\) lives.
It is then natural to identify \(L = r_0\).

Following the standard \ac{adscft} dictionary we identify the boundary value of the conjugate momentum\footnote{In terms of classical mechanics, one can think of the holographic coordinate $r$ as the ``time'' direction and then $A_t(r)$ as a generalized coordinate.} to $A_t$ with the \ac{cft} charge density
\begin{equation}
\label{Dictionary:U1CurrentIdentification}
  \frac{Q_{\acs{cft}}}{4 \pi r_0^2} = \ev{J^t} = 
  \eval{
    - \frac{r^2}{L^2} \fdv{\mathcal L_{U(1)}}{(\partial_rA_t)}}_{r\rightarrow\infty}
  =
  \frac{Q_{U(1)}}{4\pi L^2}
  ~.
\end{equation}
This tells us that the two integer parameters are naturally identified,
\begin{equation}
  \label{eq:Q-identification}
  Q_{\acs{cft}} = Q_{U(1)}.
\end{equation}
By the standard lore, we have a gravity theory with a \(U(1)\) gauge field which is  dual to a field theory with some global \(U(1)\) symmetry.
This does not contradict the fact that we have started with a non-Abelian matrix model in Section~\ref{sec:effective-action-cft}. 
As long as the ground-state physics is considered, the \ac{eft} abelianizes in the sense that the ground state at fixed \(U(1)\) charge of any matrix model %
behaves like the $O(2)$ vector model (for a justification see Section~\ref{sec:higher-order-CFT}).
The identification Eq.~\eqref{Dictionary:U1CurrentIdentification} then tells us that the bulk gauge current is naturally mapped to the \(U(1)\) current which we have fixed in Eq.~\eqref{MatrixModels:U1current}. 

Next, we compute the boundary stress-energy tensor
directly defined \cite{Balasubramanian:1999re} as the variation of the boundary action with respect to the induced boundary metric $h_{ab}$ ($a,b$ run over the boundary coordinates),
\begin{align}
  \ev{T^{a}_{\phantom{a}b}} =
  \eval{- \frac{2}{\sqrt{-h}} \pqty{\frac{r}{L}}^3 \fdv{h_{a}^{\phantom{a}b}} \pqty {S_\text{GHY} + S_\text{c.t.}}}_{r\rightarrow\infty}
	~,
\end{align}
where the boundary action terms can be found in~\eqref{BoundaryTerms:GHY} and~\eqref{AsymptoticsCoutnerTerm}.
On-shell, we find that
\begin{align}
  \ev{T^t_{\phantom{t}t}} = -\frac{M}{4\pi L^2} 
	&& \text{and} &&
	\ev{T^\theta_{\phantom{\theta}\theta}} = \ev{T^\phi_{\phantom{\phi}\phi}} = \frac{M}{8\pi L^2}
	~,
\end{align}
with $M$ the mass of the black hole in Eq.~\eqref{BlackHoleMass}.
All off-diagonal components are vanishing.
Clearly, tracelessness of the energy-momentum tensor is satisfied,
as it should for a boundary \ac{cft}.
Using the gravity-hydrodynamics dictionary (see \cite{Rangamani:2009xk} for a review)
we can associate the energy density %
and pressure $P$ of the conformal superfluid at hand, to the time-like and space-like components of the boundary tensor, respectively:%
\begin{align}
  \label{BoundaryEnergyMomentumTensor:Evaluated}
  \ev{T^t_{\phantom{t}t}} = -\frac{M}{4\pi L^2} ~= - \frac{E}{4 \pi L^2}
  && \text{and} &&
                   \ev{T^\theta_{\phantom{\theta}\theta}} = \ev{T^\phi_{\phantom{\phi}\phi}} = \frac{M}{8\pi L^2}
                   ~= P
	~.
\end{align}
In turn, this immediately implies that 
the (first) speed of sound of the conformal Goldstone is $c_s^2 := \pdv*{P}{\epsilon}= 1/2$, in agreement with Eq.~\eqref{FirstSpeedOfSound}.
We will discuss this in detail from the point of view of the gravity dual in Section~\ref{sec:quantum-fluctuations}.%

Equations~\eqref{eq:Q-identification} and~\eqref{BoundaryEnergyMomentumTensor:Evaluated} show that our natural guess is consistent with the standard \ac{adscft} identifications and we can proceed with matching the other parameters appearing in the expression for the \ac{eft} energy in Eq.~\eqref{eq:EFT-energy} and the expression for the black hole mass in Eq.~\eqref{eq:BH-mass}:
\begin{align}\label{dic1}
  \lambda &= \frac{3\pi^2}{4}\, \frac{G^2}{L^4}, &
                   b &=\frac{\sqrt{\pi G}}{e\, L},
\end{align}
or, equivalently
\begin{align}\label{dic2}
 \frac{G}{L^2} &= \frac{2}{\pi\sqrt3} \sqrt{\gl}, &
             e &= \frac{\sqrt2}{3^{1/4}} \frac{\lambda{^{1/4}}}{b}.
\end{align}

\bigskip

The usual \ac{adscft} dictionary, moreover, maps the couplings on the gravity side to quantities in the \ac{cft}.
For the \ac{adsrn} black hole, we know~\cite{Myers:2010tj,Barnes:2005bm} that 
the central charges, derived  from the two-point function of $T_{\mu\nu}$ and $J_\mu$ in the uncharged vacuum,
behave like
\begin{align}
	\label{AdS:CentralCharge}
	C_T &= \frac{3}{\pi^3}\frac{L^2}{G} ,
	& %
	C_J &= \frac{3}{\pi^2}\frac{1}{e^2} 
	~.
\end{align}

\newcolumntype{M}{>{$\displaystyle}c<{$}} %
\begin{table}
  \centering
  \begin{tabular}{MMM}
    \toprule
    \text{AdS}                       & \ac{eft}                                      & \ac{cft}                 \\
    \midrule
    \frac{L^2}{G}                    &  \frac{\sqrt{3}\pi }{2} \frac{1}{\lambda^{1/2}} & \frac{\pi^3}{3} C_T      \\[2ex]
    \frac{1}{e^2}                    & \frac{\sqrt{3}}{2} \frac{b^2}{\lambda^{1/2}}   & \frac{\pi^2}{3} C_J      \\[2ex]
    \frac{3\pi^2}{4} \frac{G^2}{L^4} & \lambda                                       & \frac{27}{4\pi^4 C_T^2 } \\[2ex]
    \frac{\sqrt{\pi G}}{e L}         & b                                             & \sqrt{\frac{C_J}{C_T}}   \\[2ex]
    \bottomrule
  \end{tabular}
  \caption{The AdS/EFT/CFT dictionary}
  \label{tab:dictionary}
\end{table}

As we already pointed out in the introduction, the central charge \(C_T\) is a measure of the total number of degrees of freedom, which scales with some model-dependent positive power of \(N\), \(C_T = \order{N^q}\).
In Eq.~\eqref{eq:EHM-action-consistency} we have seen that the \ac{adsrn} description is consistent if \( {L^2}/{G} \approx {1}/{e^2} \gg 1 \).
In terms of central charges this translates to
\begin{equation}
  \label{eq:CFT-dual-consistency}
  C_T \approx C_J =  \order{N^q} \gg 1 ,
\end{equation}
and for the \ac{eft} parameters, we find
\begin{align}
  \label{eq:EFT-dual-consistency}
  \lambda &= \order{N^{-2q}} \ll 1 , &
                                 b &= \order{1} .
\end{align}

With these identifications we see that the condition in Eq.~\eqref{Qerel} for the consistency of the gravity description becomes
\begin{equation}
  \label{Dictionary:EffectiveExpansionParameter}
  Q \gg \sqrt{C_T C_J} \approx C_T =\order{N^q}  \gg 1 .
\end{equation}
In this regime, both semiclassical descriptions in terms of \ac{eft} and black hole gravity are consistent.

The complete \textsc{ads-eft-cft} dictionary is given in Table~\ref{tab:dictionary}. In the next section we will discuss and match the higher-order corrections on the two sides.

\FloatBarrier

\section{Higher-order terms}
\label{sec:higher-order-terms}

So far, we have studied the very simple action associated to Eq.~\eqref{MatrixModels:IRaction} on the \ac{cft} side. However, the full Wilsonian action contains an infinity of terms compatible with the symmetries of the system. We therefore need to consider higher-order corrections which, however, are still controlled by the large-$Q$ expansion.

Similarly, we can also consider higher-order terms on the gravity side, which enter~\cite{Roy:2017hcp} as order-by-order corrections in Newton's constant $G$ or string length $l_s=\sqrt{\alpha'}$.  Although both corrections are possible, in what follows, we will mostly restrict ourselves to the low energy or $\alpha'\to 0$ limit to exclude massive string modes and only consider the higher-order terms arising from the small $G$ or (equivalently on the \ac{eft} side) small $1/N$ corrections. Because we already have established a hierarchy between $Q$ and $N$ in the large $Q$ limit, Eq.~\eqref{Dictionary:EffectiveExpansionParameter}, we will treat our background classically, without any ``quantum gravity'' modifications.

\subsection{Higher-order terms in the effective CFT}
\label{sec:higher-order-CFT}

We start from a simple model in which only the leading Goldstone and a massive mode appear,\footnote{This simple model was used in~\cite{Hellerman:2015nra} to describe the \(O(2)\) vector theory at the Wilson--Fisher fixed point.}
\begin{equation}
  \label{eq:sextic-action}  
 \mathcal L = \frac{1}{2} \del_\mu a \del^\mu a + \frac{b^2}{2} a^2 \del_\mu \chi \del^\mu \chi - \frac{\mathcal R}{16} a^2 - \frac{\lambda}{6} a^6 .
\end{equation}
The global \(U(1)\) symmetry acts by shifting \(\chi \to \chi + \epsilon\) and leaving \(a\) invariant.
It was shown in~\cite{Hellerman:2015nra} that at fixed \(U(1)\) charge \(Q\), the field \(a \) acquires a parametrically large \ac{vev} \(\ev{a} = v = \order{Q^{1/4}}\) which controls all the quantum corrections.
Similarly, this \ac{vev} controls most of the higher-derivative terms that are compatible with the Lorentz, \(U(1)\) symmetries and scale invariance and that are to be added to the sextic action.
The energy of the ground state at fixed charge is precisely the one given in Eq.~\eqref{eq:EFT-energy}.

Since we are at the critical point, there is no intrinsic scale to control the operators in the action.
They all are on the same footing, with coefficients that are generically of order \(\order{1}\).
The only thing that distinguishes them is their \(Q\)-scaling, which we immediately find once we observe that we are expanding around a vacuum where \(\ev{\chi(t,x)} = \mu t\)  with \( \mu = \order{Q^{1/2}}\) and \(\ev{a} = v = \order{Q^{1/4}}\).
The terms in the action~\eqref{eq:sextic-action} have non-negative \(Q\)-scaling:
\begin{equation}
  \begin{aligned}
    \del_\mu a \del^\mu a &= \order{Q^0} ,\\
    a^2 \pqty{\dot \chi}^2 &= \order{Q^{3/2}}, \\
    a^2 \pqty{\nabla \chi}^2 &= \order{Q^{1/2}}, \\
    R a^2 &= \order{Q^{1/2}} ,\\
    a^6 &= \order{Q^{3/2}} ,
  \end{aligned}
\end{equation}
while most higher-derivative terms have negative \(Q\)-scaling. Let us take for example the operator
\begin{equation}
  O_{-3/2} = \frac{\pqty{\del_\mu a \del^\mu a}^2}{a^6} .
\end{equation}
It respects all the symmetries and has scaling dimension 3 but when \(a\) is expanded around its \ac{vev} as \(a = v + \hat a\), it has \(Q\)-scaling \(-3/2\):
\begin{equation}
  O_{-3/2} = \frac{\pqty{\del_\mu \hat a \del^\mu \hat a}^2}{\pqty{v + \hat a}^6} = \order{\frac{1}{Q^{3/2}}} .
\end{equation}
This is not completely general. In fact, there is a family of higher-derivative operators that \emph{is not} \(Q\)-suppressed.\footnote{D.O. and S.R. would like to thank Martin Roček for discussions on this point.}
Take the operator
\begin{equation}
  O_0 = \frac{\del_\mu \chi \del^\mu \chi}{a^4} .
\end{equation}
It is Lorentz invariant and has both scaling and \(Q\)-dimension \(0\).
This means that in the Wilsonian action any of the terms in the sextic action can and has to be dressed with an arbitrary polynomial \(F(O_0)\) and that the effect of the dressing remains important in the large-charge limit.

What is the effect of these terms on the energy computed by the \ac{eft}?
One way of describing it is to say that the form of the large-\(Q\) expansion of the energy remains the same as for the sextic potential but all the coefficients are renormalized separately, \emph{i.e.} the energy of the ground state on a sphere of radius \(r_0\) is given by an expansion with unknown independent coefficients $c_n$,
\begin{equation}\label{eq:eft-corrterms}
   E r_0 = c_{3/2} Q^{3/2} + c_{1/2} Q^{1/2} + c_{-1/2} Q^{-1/2} + \dots
\end{equation}

A more efficient way to describe these terms consists in writing an effective action for the Goldstone field \(\chi\) alone, which will contain infinite terms with a manifest \(Q\)-scaling~\cite{Hellerman:2015nra,Monin:2016jmo}\footnote{See also~\cite{Son:2005rv} for a related approach to effective descriptions of non-relativistic
\acp{cft}.} (remember that \(\dot \chi\) scales as \(\order{Q^{1/2}}\) while \(\nabla \chi\) scales as \(\order{Q^0}\)):
\begin{equation}
  \label{eq:non-linear-action}
  \mathcal L = k_{3/2} \norm{\del \chi}^3 + k_{1/2} \norm{\del \chi}^{1/2} + k_{-1/2} \norm{\del \chi}^{-1/2} + \dots ,
\end{equation}
where \(\norm{\del \chi} = \pqty{\del_\mu \chi \del^\mu \chi}^{1/2}\).
The \ac{eft} does not give us any control over the terms that are generically of order one.
We will however see in the next section that using the \ac{adseft} dictionary we can at least estimate their dependence on \(N\).

The very same considerations apply for any other system that exhibits the symmetry-breaking pattern discussed in Section~\ref{sec:effective-action-cft}.
By construction, the action in Eq.~\eqref{eq:non-linear-action} describes the physics of the leading Goldstone field.
We come back to this point in the beginning of Section~\ref{sec:quantum-fluctuations}.

The ultimate goal of the \ac{eft} analysis is an asymptotic formula for the energy on a sphere of radius \(r_0\) which, via the state/operator correspondence, is the same as a formula for the conformal dimension of the lowest operator of charge \(Q\).
The first terms in this expansion are so far 
\begin{equation}
  \label{eq:conformal-dimension-EFT}
  D(Q) = E r_0 = c_{3/2} Q^{3/2} + c_{1/2} Q^{1/2} %
  + \order{Q^{-1/2}} ,
\end{equation}
where the \(c_{n}\) are parameters to be computed independently.
We will see in Section~\ref{sec:higher-order-comparison} how some information about these coefficients can be extracted from the dual gravity picture.

\subsection{Higher-order terms on the gravity side}
\label{sec:higher-order-gravity}

Now that we have a better understanding of the possible higher-order terms on the \ac{eft} side, it is a natural question whether this picture also appears in gravity. Indeed, the answer to this question is yes: these higher-order terms appear as higher-derivative corrections to the metric. 
Let us briefly comment on the nature of the corrections that we have in mind. For specificity, let us take the \ac{gb} correction  which is achieved by adding to the standard \ac{ehm} Lagrangian a term like 
\begin{equation}
\label{eq:GBterm}
\mathcal{L}_{\acs{gb}}=\alpha_{\acs{gb}} \ell^2 \left(\mathcal R^2 - 4 R^{\mu\nu}R_{\mu\nu} + R^{\mu\nu\rho\sigma}R_{\mu\nu\rho\sigma}\right).
\end{equation}
Here $\alpha_{\acs{gb}}$ is a dimensionless $\order{1}$ coefficient %
and $\ell$ is a length parameter which must be raised to the appropriate power for dimensional reasons. For example, the higher-order Lovelock terms, which are higher order corrections in curvature, should come with higher powers of $\ell$. The quantities $\mathcal R$, $R_{\mu\nu}$ and $R_{\mu\nu\gamma\delta}$ are respectively the usual Ricci scalar, Ricci tensor and Riemann tensor.  

The choice of the length parameter $\ell$ above indicates the bulk perturbation expansion under consideration. As mentioned at the beginning of this section, we can either consider it to be $\ell^2=G=l_p^2$ or (as is possible in string theory) string length corrections with $\ell=l_s$.
If we restrict ourselves to the $l_s\to 0$ sector, the only meaningful choice is the first one, which corresponds to perturbations in $G$ in the bulk.
Given the dictionary in Table~\ref{tab:dictionary}, in turn, we can recognize them as nothing but $1/N$ perturbations on the boundary (in case of the $O(N)$ vector models or $SU(N)$ matrix models etc.).

{At order $\ell^2$, the generic quadratic term in the curvature takes the form}
\begin{equation}\label{eq:arbitcoeff}
\begin{aligned}
S_{\text{corr}}  &=
\eval{\frac{\ell^2}{16\pi G} \int \dd[4]{x}\sqrt{-g} \left(\alpha_1\mathcal R^2 + \alpha_2 R^{\mu\nu}R_{\mu\nu} + \alpha_3 R^{\mu\nu\rho\sigma}R_{\mu\nu\rho\sigma}\right)}_{\text{finite part}} ,
\end{aligned}
\end{equation}
where $\alpha_i$ are numerical parameters.
For general values of $\alpha_i$ one might wonder if we are not changing the nature of the {\ac{eom}} thus introducing new {\acp{dof}} that would spoil the holographic correspondence.
We will take the pragmatic stance that in our description these new terms are to be seen as perturbations around the usual {\ac{ehm}} action and assume that when they are expanded around the {\ac{adsrn}} solution they do not change its structure.
Evaluating this correction, we can easily see that the large-charge expansion shows the same behavior as Eq.~\eqref{bhmass-Q},
\begin{multline}
\label{eq:curvcorrectedexp}
S_{\text{corr}}
\propto 
\frac{\ell^2}{L G}
\int \dd{t}\dd{\Omega} \Bigg[ 
\frac{30\alpha_1 + 3 \alpha_2 - 8 \alpha_3}{15} \left(\sqrt{\frac3\pi}\frac{\sqrt G}{L}e Q\right)^{3/2} \\
-
\frac{6\alpha_1 + 3 \alpha_2 + 4 \alpha_3}{2} \left(\sqrt{\frac3\pi}\frac{\sqrt G}{L}e Q\right)^{1/2}
+ \mathcal{O}(Q^{-1/2})
\Bigg].
\end{multline}
As mentioned before, if $\ell^2=G$, this contribution is suppressed by $G/L^2=\order{1/C_T}$ against the leading \textsc{eh} term. 
One special choice for the quadratic term in the curvature is the square of the Weyl tensor. 
The corresponding correction to our \ac{adsrn} background~\eqref{eq:RNsolution} has no contribution of order \(\order{Q^{1/2}}\):
\begin{align}\label{eq:weylsq}
S_\text{Weyl}  &=  
\frac{\alpha_{\text{Weyl}}}{16\pi} \frac{\ell^2}{G} \int \dd[4]{x}\sqrt{-g} \, C_{\mu\nu\rho\sigma} C^{\mu\nu\rho\sigma}
\nonumber\\
& \propto \frac{\alpha_\text{Weyl}}{L} \frac{\ell^2}{G}
\int \dd{t} \dd{\Omega} \left[ 
\left(\sqrt{\frac3\pi}\frac{\sqrt G}{L}e Q\right)^{3/2}
+ \mathcal{O}(Q^{-1/2})
\right].
\end{align}

{Another special choice is the {\ac{gb}} term corresponding to $\alpha_1=\alpha_{\acs{gb}}$, $\alpha_2=-4\alpha_{\acs{gb}}$ and $\alpha_3=\alpha_{\acs{gb}}$.
In four dimensions, it is a total derivative and does not change the {\ac{eom}}.}

\bigskip

In summary, we see that the higher-order corrections to the \ac{adsrn} solution can still be organized in a series expansion in the parameter \(\pqty{3G}^{1/2} e Q/ \pqty{\pi L}\), with leading power \(3/2\).
{Now, each coefficient in this expansion} is itself a series in the scale \(\ell\).
{The same organization of the corrections is possible for the mass of the black hole that will generically have the form}
\begin{multline}\label{eq:grav-corrterms}
  ML = \frac{L^2}{G}\Bigg[ d_{3/2} \left(\sqrt{\frac3\pi}\frac{\sqrt G}{L}e Q\right)^{3/2} + d_{1/2} \left(\sqrt{\frac3\pi}\frac{\sqrt G}{L}e Q\right)^{1/2} \\
  + d_{-1/2} \left(\sqrt{\frac3\pi}\frac{\sqrt G}{L}e Q\right)^{-1/2} + \dots \Bigg]
\end{multline}
with the expansion coefficients ($n$ counts the order in the $Q$ expansion)
\begin{equation}
  d_{n} = d_{n}^{(0)} %
  + d_{n}^{(2)} \frac{\ell^2}{L^2} + \dots ,
\end{equation}
where we have explicitly recorded the overall $L^2/G=\order{C_T}$ factor, such that all $d_n^{(i)}$ are order-one coefficients in the {$1/C_T$ expansion}.

One possible way of reading Eq.~\eqref{eq:grav-corrterms} is as the energy for an extremal \ac{adsrn} blackhole with parameters fixed by \(d_{3/2}\) and \(d_{1/2}\), plus corrections starting at order \(\order{Q^{-1/2}}\).
If we for example consider corrections up to order $\ell^2$, then the coefficient of the first $Q$-suppressed term is%
\begin{equation}
  \label{eq:renormalized-d}
d_{-1/2} = -\frac{1}{6} \frac{(d_{1/2})^2}{d_{3/2}} - 
\frac{\ell^2}{L^2}\left(\alpha_2+4\alpha_3\right) + \order*{\ell^4}	.
\end{equation}
For $\ell=0$ we recover the expression for the \ac{adsrn} black hole, where $d^{(0)}_{-1/2}$ is fixed in terms of the $d^{(0)}_{3/2}$ and $d^{(0)}_{1/2}$ coefficients
\begin{equation}
  d_{-1/2}^{(0)} = - \frac{1}{6} \frac{(d_{1/2}^{(0)})^2}{d_{3/2}^{(0)}} .
\end{equation}
However, as soon as the first correction in $\ell^2/L^2$ is incorporated, then $d_{-1/2}$ generally develops an independent, subleading piece.
For the special case of the \ac{gb} term, where $\alpha_2=-4\alpha_{\acs{gb}}$ and $\alpha_3=\alpha_{\acs{gb}}$ this correction vanishes,
showing that the \ac{gb} term preserves the nested square roots in Eq.~\eqref{eq:BH-mass}.

\subsection{Comparison}
\label{sec:higher-order-comparison}

From the discussion of the higher correction terms of both sides of the correspondence, we found Eq.~\eqref{eq:eft-corrterms} on the \ac{eft} side and Eq.~\eqref{eq:grav-corrterms} on the gravity side. It is therefore natural to identify 
\begin{equation}
  c_n = d_n \left( \sqrt{\frac3\pi}\frac{\sqrt G}{L}e \right)^n.
\end{equation}
The first observation is that the two expansions are mutually compatible.
But we learn more than that.
In fact, on the \ac{eft} side, we had no control over the expansion coefficients $c_n$.
On the gravity side, however, we have seen that the $d_n$ are given by an expansion in $\ell$, where the lowest term is the \ac{adsrn} black hole itself.
Using the dictionary in Table~\ref{tab:dictionary}, we can also write the $c_n$ as an asymptotic expansion in $1/C_T=\order{N^{-q}}$:
\begin{equation}
  c_{n} = C_T^{1-n}\left(c_{n}^{(0)} %
  + c_{n}^{(2)}\frac{1}{C_T} + \dots\right) ,
\end{equation}
where the zeroth-order coefficients $C_T^{1-n}c_{n}^{(0)}$ are obtained by $Q$-expanding the ground state energy~\eqref{eq:EFT-energy} computed from the \ac{ir} Lagrangian~\eqref{MatrixModels:IRaction}.
This suggests that the corresponding truncated action %
appropriately describes the physics in the regime of large $N$, where $Q \gg N^q$, as discussed before.

\section{Quantum fluctuations}
\label{sec:quantum-fluctuations}

\paragraph{The relativistic Goldstone and its Casimir energy.}
As already discussed in Section~\ref{sec:effective-action-cft}, the first non-trivial correction to the energy $E(Q)$ derived within our \ac{eft} is expected to appear at order $Q^0$ due to the relativistic Goldstone in Eq.~\eqref{FirstSpeedOfSound}.
The dominant effect of this gapless mode in the effective action is most easily understood by using the non-linear sigma model introduced in Eq.~\eqref{eq:non-linear-action} in terms of the $\chi$ field.
The key observation is that once expanded in the (normalized) scalar fluctuations \(\pi(t,x) = \pqty{ 6 k_{3/2} \mu}^{1/2} \pqty{ \chi(t,x) - \mu t} \),
the effective action of Eq.~\eqref{eq:non-linear-action} contains at order \(Q^0\) only the (quadratic) kinetic term for $\pi(t,x)$:
\begin{equation}
  \begin{aligned}
    k_{3/2} \norm{\del \chi}^3 &= k_{3/2} \mu^3 + \frac{1}{2} \pqty{ \pqty{\del_t \pi }^2 - \frac{1}{2} \pqty{\nabla \pi}^2} + \order{\frac{1}{\mu^{3/2}}} , \\
    k_{1/2} \norm{\del \chi} &= k_{1/2} \mu + \order{\frac{1}{\mu^{1/2}}}.
\end{aligned}
\end{equation}

The first line gives us precisely the anticipated kinetic term for a relativistic Goldstone with speed of sound \(c_s = 1/ \sqrt{2}\).
At this stage, we recall that neither the ground-state solution nor any higher corrections (discussed in Section~\ref{sec:higher-order-CFT}) contribute at $\mathcal{O}(Q^0)\sim \mathcal{O}(1)$ in Eq.~\eqref{eq:conformal-dimension-EFT}.
This automatically means that the only source of terms with \(Q^0\)-scaling in the energy expansion are quantum fluctuations, more precisely, the Casimir energy \cite{Monin:2016bwf} of the field \(\pi(t,x)\):
\begin{equation}
\label{eq:CFT:CasimirEnergyS2}
E^\pi_\text{Casimir}(r_0) = 
\frac{c_s}{2r_0} \sum_{l\in\setN} (2l+1)\sqrt{l(l+1)}
\simeq \frac{c_s}{2r_0}\left(-\frac{1}{4} - 0.015096\right) \simeq -0.094 \times \frac{1}{r_0}.
\end{equation}
This Casimir energy modifies our prediction for the anomalous dimension Eq.~\eqref{eq:conformal-dimension-EFT} such that the final result of the \ac{eft} analysis becomes 
\begin{equation}
  \label{eq:conformal-dimension-EFT-final}
  D(Q) = E r_0 = c_{3/2} Q^{3/2} + c_{1/2} Q^{1/2} - 0.094 + \order{Q^{-1/2}} .
\end{equation}
In the following paragraph %
we will find a precise agreement between this analysis and the spectrum of metric and gauge-field fluctuations in the dual picture.

\paragraph{Speed of sound in the gravity picture.}

In Section~\ref{sec:linear-sigma-model-and-black-hole} we have observed that the leading quantum correction in the \ac{eft} comes from a relativistic Goldstone mode with speed of sound \(c_s^2 = 1/2\).
This value can also be derived from general thermodynamic arguments (see the comment %
below Eq.~\eqref{BoundaryEnergyMomentumTensor:Evaluated}).
Moreover, we have seen that the corresponding contribution to the energy is classically protected, in the sense that there are no corrections of order \(Q^0\) coming from the higher-order terms either in the \ac{eft} or in the gravity picture.

It is then natural to expect this mode to appear also in the gravity dual.
This is in fact the case, as it was shown in~\cite{Edalati:2010pn,Edalati:2010hk}.
We summarize here their argument.
The idea is to study the fluctuations of the \ac{ehm} action around the \ac{adsrn} black hole solution of Eq.~\eqref{eq:RNsolution}.
The authors of~\cite{Edalati:2010pn,Edalati:2010hk} then write the metric and gauge field as
\begin{align}
  g = g^{\acs{rn}} + \hat g && \text{and } && A = A^{\acs{rn}} + \hat A
\end{align}
and choose the radial gauge
\begin{equation}
\hat g_{tr} = \hat g_{rr} = \hat g_{\theta r} = \hat g_{\phi r} = 0 
\qand
\hat A_r = 0.
\end{equation}
After Fourier-transforming the non-vanishing modes in momentum-space with $p_\mu=(\omega,\vec p)$ along the boundary $\mathbb R_t \times S^2$, they look for solutions to the linearized (in $\hat g$ and $\hat A$)  \ac{eom} at a generic point in $r\geq r_h$.
The investigation is exhaustively carried out combining analytic manipulations of the coupled system of linearized equations and ultimately numerical simulations, working in the hydrodynamic regime of small momentum $\vec p \ll \mu$. Given that \(\mu = \order{Q^{1/2}}\), this hydrodynamic limit in our language corresponds to the large-charge scenario described by Eq.~\eqref{eq:EFT_ValidityRegime}. 
The final result of the analysis
of the quasinormal modes of the \ac{adsrn} black hole is that the corresponding boundary \textsc{cft} includes
eight massive modes with gap proportional to the chemical potential \(\mu = \order{Q^{1/2}}\) and one relativistic Goldstone with dispersion
\begin{equation}
  \label{eq:gravity-Goldstone}
  \omega(p) = 0.704 \, p \approx \frac{1}{\sqrt{2}} p.
\end{equation}

This spectrum reproduces the leading part of the \ac{eft} analysis of Section~\ref{sec:linear-sigma-model-and-black-hole}, including the massive modes with mass \(\order{Q^{1/2}}\) discussed in~\cite{Loukas:2017lof,Loukas:2017hic}.
The fact that any gapped mode coming from the fluctuations around the \ac{adsrn} background appears with parametrically large mass (as anticipated from the \ac{eft} picture) serves as a verification for the self-consistency of the gravity description.
  In spite of this nice confirmation of our observation and conjecture, we have to point out that the spectrum of the quasinormal modes contains subleading (in \(Q\)) imaginary terms, \emph{i.e.} a non-vanishing attenuation constant.
  This is conspicuously absent in the spectrum of the \ac{eft}.

What is manifestly absent though, are the non-relativistic gapless modes of Eq.~\eqref{eq:non-relativistic-Goldstones}. This shows that the simple \ac{adsrn} black hole cannot describe the full symmetry-breaking pattern that characterizes a \ac{cft} with non-Abelian symmetry group at fixed (and large) charge.
This seems to imply that the symmetry breaking pattern relevant for the \ac{adsrn} black hole is the most minimalistic described by the breaking of exactly one $U(1)$ subgroup.
Despite this, we have seen that already our simple black hole solution can reproduce the dominating effects up to order one in the large-charge expansion of the anomalous dimension in Eq.~\eqref{eq:conformal-dimension-EFT}.

\paragraph{Higher-order terms.}

The computation that we have outlined above is performed in pure \ac{ehm} gravity.
How is it influenced by higher-order terms?
By studying the energy on the bulk side, in Section~\ref{sec:higher-order-gravity} we have already shown that such terms generically do not change the nature of the large-$Q$ expansion~\eqref{bhmass-Q}, however they do spoil the closed form structure of~\eqref{eq:BH-mass}.
On the other hand, in the general expansion there are only two terms with positive \(Q\)-scaling, which means that the leading effects of any higher-order correction can always be reabsorbed in a renormalization of the two parameters of the \ac{adsrn} black hole, thus leaving only corrections starting at order \(Q^{-1/2}\),  as we have seen in Eq.~\eqref{eq:renormalized-d}.
Since the computation of the Goldstone mode in Eq.~\eqref{eq:gravity-Goldstone} depends only on the form of the \ac{adsrn} solution, we conclude that  adding higher-order terms can only modify terms of order \(Q^{-1/2}\) and below, thus leaving the \(Q^0\) leading-order Casimir energy in~\eqref{eq:conformal-dimension-EFT-final} unaffected. %

\section{Discussion}
\label{sec:discussion}

In this paper we have discussed the {\ac{adscft}} correspondence for conformal field theories with (non)-Abelian global symmetry where a $U(1)$ charge $Q$ is held fixed 
    under the hierarchy $Q \gg C_T \gg 1$, %
    $C_T$ being the central charge.
    In this regime, both sides of the correspondence are semiclassical and become perturbatively accessible.
  We start from the observation that the ground state energy as a function of $Q$ for a simple {\ac{eft}} coincides with the expression for the mass of an {\ac{adsrn}} black hole as function of its charge.
  Building from this, we propose a dictionary relating the {\acl{eft}} description with the gravity dual, which gives us a new handle on the coefficients of the {\ac{eft}}.
  Using our gravity picture, we can promote these coefficients from simple numerical factors to actual functions of the parameters of the underlying {\ac{cft}}, such as the central charge.%

\bigskip

We would like to conclude this note with some observations relating our findings to some points raised in the literature as well as some natural directions for future investigation.

\subsection{Stability and \acl{wgc}}

An immediate question that arises in our above analysis, is whether the extremal black holes considered in AdS space give rise to a stable solution at all. 
In what follows, we will focus in particular on the superradiant instability. We will not consider the precise numerical factors, but rather will focus only on all the physical variables and parameters involved.

In~\cite{Nakayama:2015hga}, it was found that the \ac{adsrn} black hole \emph{has a superradiant instability} if there is an operator with dimension \(\Delta\) and charge \(q \) such that
\begin{equation}\label{wgcads}
\frac{\Delta^2}{q^2} < \frac{C_T}{C_J} .
\end{equation}
We are interested in the consistency of the \ac{eft} picture, which means that we look only at the sector where all operators have charge \(q = Q\) and by construction dimension \(\Delta \geq \Delta_{\acs{rn}}\).
We want to see if within this sector the superradiant instability is triggered.
One can straightaway see that with our dictionary for $C_{T}$ and $C_{J}$~\eqref{AdS:CentralCharge} and  using the leading large-$Q$ behavior of an extremal \ac{adsrn} black hole~\eqref{bhmass-Q}, we precisely retrieve the regime of validity of our approximation (using ~\eqref{eq:EHM-action-consistency}),
\begin{equation}\label{ourlimits}
  \frac{\Delta^2}{Q^2} \geq \frac{\Delta_{\acs{rn}}^2}{Q^2}> \frac{C_T}{C_J}\qquad \implies \qquad Q\gg\frac{L^2}{G} (\approx C_T).
\end{equation}
Thus we see that our regime of validity in the large $Q$ sector~\eqref{Qerel}, is consistent with the proposed version of stability of the corresponding extremal \ac{adsrn} black hole, at least with respect to this mode.
\bigskip

In this spirit, our present work also has interesting consequences for the \ac{wgc}~\cite{ArkaniHamed:2006dz}, which was also discussed in this context by~\cite{Nakayama:2015hga}. In its generality, the \ac{wgc} is a conjectured statement about a \emph{consistent} theory of quantum gravity in asymptotically flat spacetimes. It points out that for a given (dimensionless) gauge coupling $g$, there must be an associated light enough particle of mass $m$ such that (recall that $m_p^2=1/G$ is the Planck mass in four spacetime dimensions)
\begin{equation}\label{wgcstate}
m<g m_{p}.
\end{equation}
In the original paper~\cite{ArkaniHamed:2006dz}, one of the motivating statements for the \ac{wgc} was that for a black hole (with the charge $q$ playing the role of the coupling $g$), the above coupling to mass ratio must be larger than the corresponding value of an extremal black hole. In flat spacetimes, \emph{e.g.} for a \ac{rn} black hole, the extremality condition is given by 
\begin{equation}\label{extrnflat}
\frac{q_{ext}}{m_{ext}}=\frac{1}{m_p}.
\end{equation}
This then directly implies~\eqref{wgcstate}.

However, for our \acl{adsrn} black holes, the connection between the extremal mass and charge is different \emph{in the large charge sector}~\eqref{bhmass-Q}, so a similar argument must yield a different \ac{wgc} constraint. 
To understand this better, once again noting that for our extremal \ac{adsrn} black holes
\begin{equation}
{M_{ext}L}=\frac{\sqrt{L}(eQ)^{3/2}}{G^{1/4}}
\end{equation}
to leading order in the large-$Q$ expansion (and using~\eqref{eq:EHM-action-consistency}), one might be tempted to instead suggest the following charge (coupling) $q$ to mass ($m$) ratio as a potential statement for the \ac{wgc}:
\begin{equation}\label{wgcnewstate}
\frac{q^{3/2}}{m}>\frac{L^2}{\sqrt{G_{Q}}}.
\end{equation}
This is of course motivated along the comments we made between~\eqref{wgcstate} and~\eqref{extrnflat} and is clearly along the lines of~\cite{ArkaniHamed:2006dz}, where the extremal black holes that we have, can now also decay without producing remnants. 

In above we have distinguished the Newton's constant $G$ in~\eqref{wgcnewstate} from the ordinary Newton's constant so that we can understand the effect of our analysis at large charge towards the issues of stability and the \ac{wgc}. Indeed we find that for (note that $\frac{Q}{C_T}$ is also our effective expansion parameter on both the \ac{eft} and gravity side)
\begin{equation}
G_Q=G\frac{Q}{C_T},
\end{equation}
the appropriately adjusted form of~\eqref{wgcnewstate} precisely reproduces the stability statement of~\cite{Nakayama:2015hga} given in~\eqref{wgcads}. Since we are in the regime~\eqref{ourlimits}, it guarantees that we are always below the usual four-dimensional Planck scale, $G_Q^{-1/2}<G^{-1/2}=m_p$, and as long as this is true, there is a sense in which our stable solutions can simultaneously satisfy the \ac{wgc}.

\subsection{Further directions}

In the following, we collect some of the most obvious questions to consider next and some related work in progress.
\paragraph{Adding scalar hair.} %
In this article, %
we have seen that for the \ac{adseft} correspondence to work, there is no need to introduce additional ingredients beyond the metric and Maxwell field in the bulk. In other words, the fluctuations around the charged black hole alone, already include a Goldstone signaling a spontaneous breaking of $U(1)$ symmetry~\cite{Edalati:2009bi}. This is precisely what gives rise to the previously mentioned quasinormal spectrum, which can also be alternatively studied using a \emph{probe} scalar hair in the black hole background. Indeed, starting from~\cite{Gubser:2008px}, many authors have extensively investigated the possibility of a hairy (non)-extremal \ac{adsrn} black hole, especially in conjunction with holographic superconductors~\cite{Hartnoll:2008kx}. How to incorporate similar effects (with arbitrary fields on top of the gravitational background) into our large-charge description and understand their mapping to the dual \ac{eft} in a unified manner remains an open question.

\paragraph{Entropy.}

The entropy of the black hole discussed in Section~\ref{sec:reissn-nordstr-BH} is
\begin{equation}
\label{eq:BH-Entropy}
S = \frac{4\pi r_h^2}{G} = \frac{2\pi}{3} \frac{L^2}{G} \left(\sqrt{\frac{3e^2GQ^2}{\pi L^2}+1}-1\right)
=
\frac{2\sqrt\pi}{\sqrt3}\frac{e L}{\sqrt G} Q - \frac{2\pi}{3}\frac{L^2}{G} + \order{Q^{-1}} .
\end{equation}
Using the dictionary of Section~\ref{sec:dictionary}, this can be rewritten as
\begin{equation}
  \label{eq:BH-entropy-QN}
  S = s_1 Q + s_2 C_T + \order{Q^{-1}},
\end{equation}
where \(s_1\) and \(s_2\) are constants of order \(1\).

It is a natural question to ask whether it is possible to reproduce this behavior from the \ac{eft} at zero temperature. Taking into account finite temperature effects may shed some light on this issue. In particular, purely from the perspective of a Wilsonian effective action, we expect the boundary theory to have a finite entropy at finite temperature. This must be reflected in any potential bulk dual. %
We defer the resolution of the entropy of the system using \ac{eft} techniques to future work.

\paragraph{Near-horizon $AdS_2$ and \(su(1,1)\).}

In~\cite{Loukas:2016ckj} the field theory of a complex scalar was investigated at large charge. 
In particular, the associated $0+1$ dimensional vacuum was explicitly constructed as a (generalized) coherent state showing the emergence of the $su(1,1)$ algebra as the operator algebra governing the quantum mechanics at fixed (and large) charge.
The same algebra appears naturally~\cite{Claus:1998ts,Kallosh:1999mi} as an isometry of the near-horizon limit ($r_\epsilon=r-r_h$) of the extremal black hole, which is the Bertotti--Robinson black hole \(AdS_2 \times S^2\) with metric
\begin{equation}
  \label{NearHorizonGeometry}
  \dd{s^2} = 
-\frac{6r_\epsilon^2}{L^2}\dd{ t^2} + \frac{L^2}{6r_\ge^2}\dd{r^2_\epsilon} 
+ \frac{\sqrt G}{L} e Q \frac{L^2}{\sqrt{12 \pi}} \left( \dd{\theta^2} + \sin^2\theta\dd{\phi^2}\right) .
\end{equation}
This observation might help to shed some more light on the %
conformal quantum mechanics dual to the \ac{ir} geometry of the extremal black hole. %

\paragraph{The vector models.}

The \acl{eft} discussed in Section~\ref{sec:linear-sigma-model-and-black-hole} was originally introduced to describe the Wilson--Fisher point of \(O(N)\) vector models~\cite{Hellerman:2015nra,Alvarez-Gaume:2016vff}.
These models do not have a known simple gravity dual, but one can wonder if they can still be related to the \ac{adsrn} black hole of Section~\ref{sec:reissn-nordstr-BH}, at least in the limit of large charge.

We would like to observe that if we try to apply the dictionary in Table~\ref{tab:dictionary} and use the results of~\cite{Osborn:1993cr,Petkou:1995vu} for the central charges \(C_T\) and \(C_J\) we still find results that are consistent with the gravity analysis.
In particular we see that \(L/\sqrt{G} = \order{N^{1/2}}\) and \(e = \order{N^{-1/2}}\), in agreement with the requirements in Eq.~\eqref{eq:EHM-action-consistency}.
It would be interesting to understand if this observation can be pushed further and possibly compare holographic computations with the lattice results at fixed charge such as the ones in~\cite{Banerjee:2017fcx}.

\section*{Acknowledgments}
The authors would like to acknowledge illuminating discussions with Matthias Blau, Simeon Hellerman, Yu Nakayama, Martin Roček and Konstantinos Siampos, and would like to thank Matthias Blau for detailed comments on the manuscript.

The work of O.L. and S.R. is supported by the Swiss National Science Foundation (\textsc{snf}) under grant number \textsc{pp00p2\_157571/1}.
D.O. acknowledges partial support  by \textsc{nccr 51nf40-141869} ``The Mathematics of Physics'' (SwissMAP). D.S. is supported through \textsc{nccr 51nf40-141869} ``The Mathematics of Physics'' (SwissMAP).

\appendix

\section{The free energy of the boundary theory} \label{sec:app}

Here we wish to compute the free energy of the boundary theory at $r\rightarrow\infty$. 
Since the bulk action evaluated at the classical solution is naively divergent, we first need to perform the standard regularization of gravity divergences.
Therefore, we introduce the regularized action 
\begin{equation}
S_\text{reg} = \left(S_\text{EH} + S_{U(1)}\right) + \left(S_\text{GHY} + S_{c.t.}\right)
~,
\end{equation}
in terms of the original bulk action and additionally the Gibbons--Hawking--York boundary term (to ensure a well-defined variational problem)
\equ{
	\label{BoundaryTerms:GHY}
	S_\text{GHY} = \frac{1}{8\pi G} \int \dd[3]x \sqrt{-h} \,K %
	~,
}
and the counter-term action~\cite{Emparan:1999pm} suitable for four-dimensional AdS spaces,
\begin{equation}
\label{AsymptoticsCoutnerTerm}
S_{\text{c.t.}} = -\frac{1}{8\pi G} \int \dd[3]x \sqrt{-h} \left(\frac{2}{L} + \frac{L}{2}  R_\infty %
\right)
.
\end{equation}
$h$ denotes the induced metric at the boundary,
$K=h^{ab} K_{ab}$ is the trace of the extrinsic curvature $K_{ab}$ ($a,b$ run over the boundary coordinates) and $R_\infty$ the Ricci scalar at the boundary hypersurface. 

We now apply the main \ac{adscft} correspondence statement for the free energy
of the dual \ac{cft}, meaning we compute the {on-shell} action $S_\text{reg}$ (at any temperature $T$):
\begin{align}
F_\text{cl}   %
=
T\, S_\text{reg}[g_\text{cl},A_\text{cl}]
=
\frac{1}{4G} \left(\frac{e^2G}{4\pi r_h} Q^2 {\,-\,} r_h  + \frac{r_h^3}{L^2} \right)
\, =\, \frac12 \left(M -\frac{r_h}{G} \right)
,
\end{align}
in terms of the black hole mass in Eq.~\eqref{BlackHoleMass}.
Following~\cite{Cai:2001jc} we identify the term subtracted from $M$ as the ``Casimir energy'' $E_c$ {(not to be confused with the Casimir energy of the Goldstone as discussed in Section~\ref{sec:quantum-fluctuations})}
due to the spherical part in our global $\text{AdS}_4$ metric.
At extremality, $T=0$, we see that $E_c =  \frac{r_h}{G} \sim \order{\sqrt Q}$. Currently we are working on a detailed thermodynamical map between the two sides of the duality at finite $T$.

\setstretch{1}

\printbibliography

\end{document}